# High-Speed, Photon Counting CCD Cameras for Astronomy

Craig Mackay[a,*], Tim D. Staley[a], David King[a], Frank Suess[a] and Keith Weller[a]
[a]Institute of Astronomy, University of Cambridge, Madingley Road, Cambridge, CB3 0HA, UK;

## ABSTRACT

The design of electron multiplying CCD cameras require a very different approach from that appropriate for slow scan CCD operation. This paper describes the main problems in using electron multiplying CCDs for high-speed, photon counting applications in astronomy and how these may be substantially overcome. With careful design it is possible to operate the E2V Technologies L3CCDs at rates well in excess of that claimed by the manufacturer, and that levels of clock induced charge dramatically lower than those experienced with commercial cameras that need to operate at unity gain. Measurements of the performance of the E2V Technologies CCD201 operating at 26 MHz will be presented together with a guide to the effective reduction of clock induced charge levels. Examples of astronomical results obtained with our cameras are presented.

## 1. INTRODUCTION

For many years, astronomers have been aware of rapidly varying emission from a variety of objects. Much of the data obtained so far has been obtained at x-ray and gamma-ray wavelengths as well as at radio wavelengths. Highly variable objects such as gamma ray bursters, neutron star and black hole binaries, pulsars and cataclysmic variables are all providing important insights into the mechanisms by which their radiation is produced. There is potentially a great deal of information that could be had in the visible and near infrared bands. Although visible band photon counting detectors have been available for many years (photomultipliers and avalanche photodiodes as single element detectors, and intensified photon counting detectors for imaging applications) they have generally had problems that significantly limit their application. Recently, an exciting new technique known as Lucky Imaging has been developed that has already delivered the highest angular resolution images ever taken in the visible or near infrared. It relies heavily on the availability of high-speed, photon-counting imaging detectors that are fast enough to freeze the image motion experienced on ground-based telescopes generated by atmospheric turbulence.

Charge coupled devices (CCDs) are now ubiquitous for astronomical imaging but their characteristically high readout noise at high frame rates has limited their application in high time resolution astronomy. Recently, however, a new CCD architecture has been developed by E2V Technologies Ltd. (Chelmsford, UK) which offer a virtually noiseless amplification mechanism while retaining all the advantages of standard CCD technology such as very high quantum efficiency, excellent charge transfer efficiency and superb image quality. These electron multiplying CCDs (EMCCDs) are now available commercially both from E2V Technologies Ltd and from Texas Instruments (Dallas, US).

This paper will discuss in some detail the capabilities and limitations of this new technology as it may be applied to astronomy applications. The paper only discusses the use of those devices manufactured by E2V technologies Ltd but it is expected that many of the general considerations will be relevant to devices made by Texas Instruments. Such systems undoubtedly have the capability of transforming high time resolution imaging provided the greatest care is taken in the design, testing and optimisation.

---

* Craig Mackay: cdm <at> ast.cam.ac.uk

## 2. ELECTRON MULTIPLYING CCD (EMCCD) TECHNOLOGY

Charge coupled devices (CCDs) are silicon integrated circuits which excite an electron from a bound state when a photon is absorbed. The electrons are held in position by an array of electrodes deposited on one surface of the CCD. As the exposure progresses an electronic equivalent of the scene being imaged is accumulated. At the end of the exposure the voltages on the electrodes are clocked to transfer the entire image to one edge of the device. Each row is transferred into a readout register. Electrons in the elements of the readout register are similarly transferred to the end of the register by clocking the register electrodes. A buffer transistor matches the output to the signal processing electronics that are normally off-chip. In standard CCDs this buffer amplifier has a finite read noise (the noise level it produces in the absence of any input signal) that can range from a few electrons RMS at pixel rates of hundreds of kilohertz up to several hundred electrons at tens of megahertz pixel rate. For high-speed applications this readout noise is a significant limitation on the minimum scene illumination that the system can work with. The electron multiplying CCD technology developed by E2V Technology Ltd[1] extends the output register (figure 1).

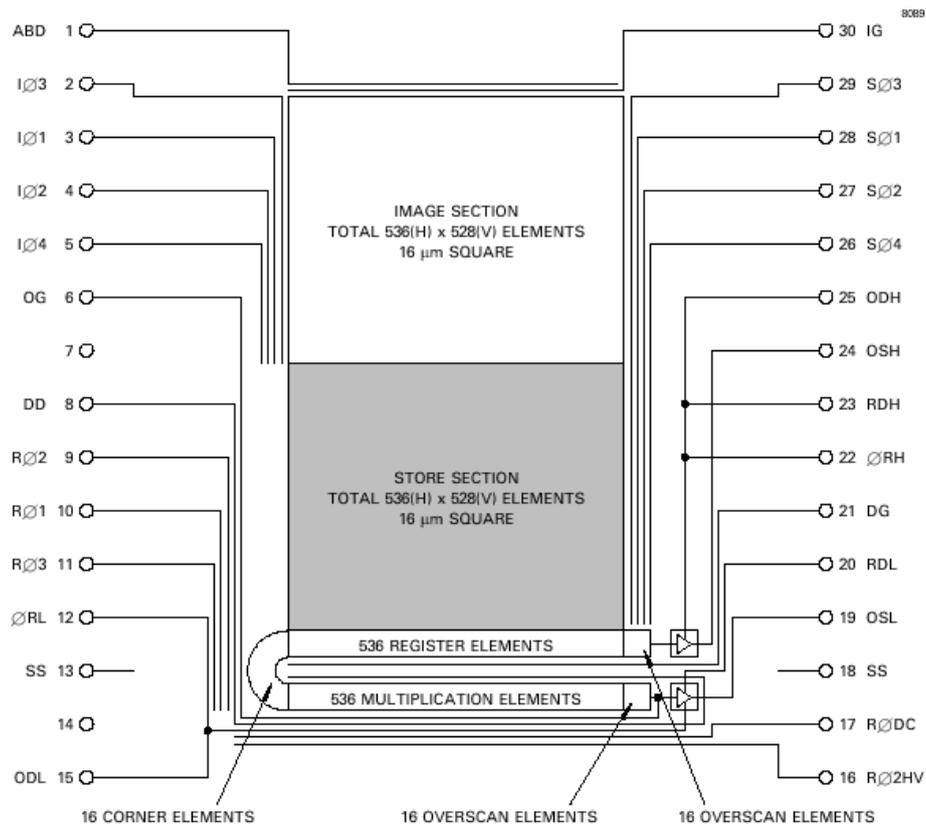

Figure 1: The layout of the E2V Technologies Ltd CCD 201 electron multiplying CCD. The output register has a 604 multiplication elements.

Within the extended (multiplication) register one of the sets of electrodes is operated at a much higher voltage swing, typically ~ 40 V compared with ~12 V for both parallel (image) and serial registers . By careful control of the conditions under these high voltage clock lines there is a small possibility of avalanche multiplication, typically in the region of ~1 percent per transfer. A multiplication register of ~600 elements will then give a gain of ~390 with a 1% avalanche probability and ~7600 with probability of 1.5%. The gain level is selected by adjusting the high-voltage clock swing.

The properties of CCDs are now remarkably good. Back illumination (thinning) routinely delivers peak quantum efficiencies in excess of 95%, charge transfer efficiency is extremely good and the cosmetic quality of the devices is remarkable so that modern CCDs are often blemish free. Electron multiplying CCDs have the same qualities but the multiplication register allows an essentially noiseless amplification of the electron signal produced by the device. The effect of the gain is that the readout noise of a CCD with no multiplication is effectively divided by the multiplication gain. With gains of hundreds or thousands the equivalent readout noise that can be achieved will be very much less than one electron RMS. Provided the CCD is able to transfer charge reliably there is no reason why single electrons may not be detected with good signal-to-noise, even at very high pixel readout rates.

The ideal photon counting detector would be able to handle images at very low signal levels, levels very much lower than normally encountered in CCD imaging. Characteristics of CCDs that are normally well below the level of detection may become highly significant and limit what electron multiplying CCDs are capable of achieving. In Cambridge we are engaged in the development of high-speed imaging systems for use in astronomy principally for Lucky Imaging and for use with curvature wavefront sensors. In both these applications signal levels can be very low indeed, particularly when the detectors are operated at the highest frame rates possible. This paper reports the results of these developments in order to give a greater insight into how the best results may be obtained and what the limitations are likely to be when other applications are considered.

## 3. ELECTRON MULTIPLICATION STATISTICS

The stochastic nature of the multiplication process is such that for any gain level with a single input electron the most probable output is still one electron. The average output is the gain level in electrons and the probability distribution of the output level is a monotonically decreasing function. This produces a substantial variance in the signal that was not present before multiplication. This is described in some detail by Basden et al[2] where it is shown that the noise added by this multiplication is equivalent to halving the detective quantum efficiency of the device. For a signal of N electrons per pixel for which we would expect a shot noise of $\sqrt{N}$ whereas we in fact find a noise equal to $\sqrt{2N}$. However when we are working at signal levels well below one photon per pixel per frame then, provided the gain is high enough to give a good signal-to-noise on each photon, the image may be processed to replace events of adequate signal-to-noise with a single constant value. This eliminates the added variance although it does mean that there will be real events that are excluded by the process if they happened to be below the threshold. Reducing the threshold increases the chance of spurious events such as conventional read-out noise outliers being detected. A threshold of ~4-5 sigma on the read-out noise largely eliminates the effect of spurious events and, with a suitably high gain, will only lose a few percent of detected events.

Basden et al[2] also examined the situation when signals of a few electrons per pixel per frame are being detected. By looking at the output signal probabilities of a small number N of input electrons to the multiplication register they showed that one could make a reasonably intelligent guess about the signal levels detected so as to replace the variable signals with 1, 2, 3, etc. This reduced the variance significantly at the lowest values for N but had essentially no effect once N > 5. In the work reported below we will concerned principally with signal levels of below one event per pixel per frame where a simple thresholding eliminates the multiplication variance.

## 4. DARK SIGNAL AND CLOCK INDUCED CHARGE

When the CCD is in total darkness, a background signal will gradually accumulated as a consequence of the ambient temperature of the device. This thermally generated dark signal has two components, the surface dark signal and the bulk dark signal. The dark signal accumulates linearly with time inside the device. The characteristics of these dark signals are described in the "Low-Light Technical Note Number 4" from E2V Technologies Ltd.[3] The surface dark current is typically ~100 times that of the bulk dark signal. By operating the CCD in inverted mode the surface dark current can be substantially suppressed. In the astronomical context it is often required to use very substantial amounts of cooling in order to suppress the dark current from either source. CCDs manufactured by E2V Technologies are operated typically at temperatures between -100°C and -140°C. The only slightly negative consequence of deep cooling with CCDs is the reduction in sensitivity at wavelengths greater than ~800nm.

There is a second source of spurious signals within CCDs that is normally masked by the relatively high readout noise of these devices, but is very apparent with electron multiplying CCDs working with significant internal gain. Clock induced charge (CIC) is generated as a consequence of operating the clocks necessary to move the electronic image across the device. This spurious signal is only generated during readout and is therefore independent of integration or exposure time. There are many operational parameters that can affect CIC including clock amplitudes and waveforms as well as bias conditions.

It is important to think carefully about the way an electron multiplying CCD is to be operated. The manufacturers suggest (entirely reasonably) bias levels and clock amplitudes and waveforms that allow the full performance of the CCD to be realised even when it is operated in unity gain (when the high-voltage clock amplitude is too low to give any measurable gain). The relatively high surface dark signal is suppressed effectively by operating the device in inversion. However this has the effect of increasing the clock induced charge by a factor of 10-50. The clock amplitudes needed to transfer the CCD full well efficiently are significantly higher than is needed to transfer a signal with a peak of only a few electrons per pixel per frame as will be encountered in most astronomical contexts. Provided one is prepared to accept a substantially reduced full well capacity then the parallel register clock amplitudes may be reduced. With suitable biasing to maintain the surface of the CCD completely out of inversion and by using minimal parallel clock amplitudes CIC may be reduced dramatically.

Relatively small increases in parallel clock amplitude can increase CIC quite considerably. Increasing the parallel clock swings from 12.5 V to 16.5 V will increase parallel clock CIC by an order of magnitude. With careful setup, parallel clock amplitudes of 10-11 V will give excellent charge transfer efficiency with low signal levels. For this reason it is important to pay particularly careful attention to the design of the parallel clocking scheme. Astronomers traditionally use electrical connectors on the side of a liquid nitrogen cooled vacuum dewar to connect to the cooled CCD. This connector is usually wired through to the CCD using fine gauge Teflon coated wires. This has two unsatisfactory consequences when operating CCDs at high pixel rates. Firstly the overall conductor lengths become rather large making it harder to generate the fast clock pulses that are need. Secondly it becomes very difficult to match the impedances of the driver electronics to the interface cabling and then to the CCD itself. Mismatches cause overshoot and ringing on the waveforms which momentarily give much higher voltages on the clocking electrodes that can worsen the clock induced charge levels markedly. Oscilloscopes connected to the drive circuitry can give a highly misleading impression of the CCD waveforms experienced within the detector, particularly in the middle of the CCD image and store areas.

With careful design it has proved possible to reduce parallel clock induced charge generation to very small levels indeed, better than one event per ten thousand pixels per frame. The output register, however, when operated with relatively high gain levels quickly approaches the full well capacity of the register. This reduces the possible range of adjustment in clock waveform amplitudes and therefore the greatest possible care needs to be taken with the driving and matching of the serial clock lines if serial clock induced charge is to be minimised.

## 5. SYSTEM DESIGN TO MINIMISE CLOCK INDUCED CHARGE

The suggested maximum pixel readout rate for E2V Technologies electron multiplying CCDs are 15 MHz for the CCD97 (512 x 512 pixels of 16 μ square) and 20 MHz for the CCD 201 (1024 x 1024 pixels of 13 μ square). This is set because of the settling time of the output amplifier on the CCD. By placing an additional buffer transistor as close to the output source of the CCD amplifier, the load capacitance can be minimised and significantly higher readout rates may be achieved. Most of the experiments described here use a pixel rate of 26 MHz with the CCD 201, a rate limited by the electronics in our controller rather than by the CCD itself. Indeed our experiments suggest that with care operation at up to ~40 MHz should be quite practical.

If indeed it is wished to work at such high pixel rates it is important to appreciate that the electronic design of an effective controller requires very different methods to those used traditionally for astronomers at much slower readout rates. Engineers are fortunate now to have a wide range of extremely integrated circuit devices intended for the domestic digital still camera market. Here the increasing trend for greater pixel count and the ability to take many frames per second means that parts designed to allow pixel rates in the 30-100 MHz range are becoming increasingly common. At these frequencies it is essential that the track length between driver and the CCD is not only minimised but designed with carefully controlled impedances that are matched between the CCD clock electrodes and the detector on its header. We have developed a novel approach to this by using a rigid multilayer printed circuit board on which the CCD is mounted inside the dewar. The PCB is sandwiched between the body of the liquid nitrogen dewar and the front plate with "O" rings. The tracks that carry signals between the driver electronics and the CCD are taken through a wall of the dewar on inner layers of the PCB while the outer layers are left with plain copper that is gold plated (figure 2).

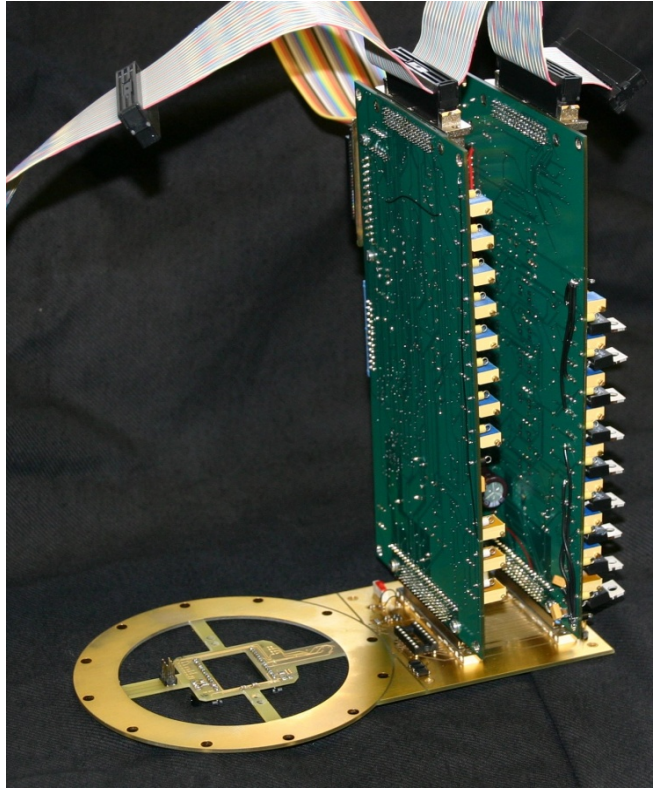

Figure 1: The basic layout of the Cambridge EMCCD controller is shown here. The headboard on which the CCD is mounted (downward looking in this image) is integrated into the vacuum dewar while the two boards provide clocking, bias and signal processing. The common data bus to the sequencer is on the edge of the boards. The power distribution bus close to it on the far edge.

By containing all the tracks in a rigid structure the impedances may be carefully controlled. The CCD is held on fibreglass spider structures to provide a mount with a small degree of flexure. The CCD sits on the metal block which is connected to the cold liquid nitrogen source by flexible braid. The metal block has temperature sensors and a heater mounted on it to allow the temperature of the CCD to be controlled accurately. This is very important because the gain of the electron multiplying CCD at a particular voltage varies rapidly with temperature in the sense that the gain is reduced as the temperature rises.

In practice, it is helpful to have adjustable series resistors in line with both low speed and high-speed clock driving circuitry in order to minimise overshoot on the clock waveforms since it does appear that this is a critical part of minimising clock induced charge levels. The high-voltage clock is particularly demanding since the evidence we have is that it is the serial clock induced charge that dominates and that with careful design the parallel clock induced charge may be reduced to essentially negligible levels. The high-voltage clock circuitry has to switch between two levels that are approximately 40 V apart in a few nanoseconds, something that is very difficult to manage while maintaining control of the waveform.

# 6. QUANTIFYING CLOCK INDUCED CHARGE

With careful design, clock induced charge generation may be reduced to very low levels indeed making its measurement difficult. It is important to develop a test strategy that allows a clear separation between clock induced charge from the parallel registers and that from the serial register (particularly the gain register). One easy way to do this is to overscan the CCD in the output register direction (figure 3).

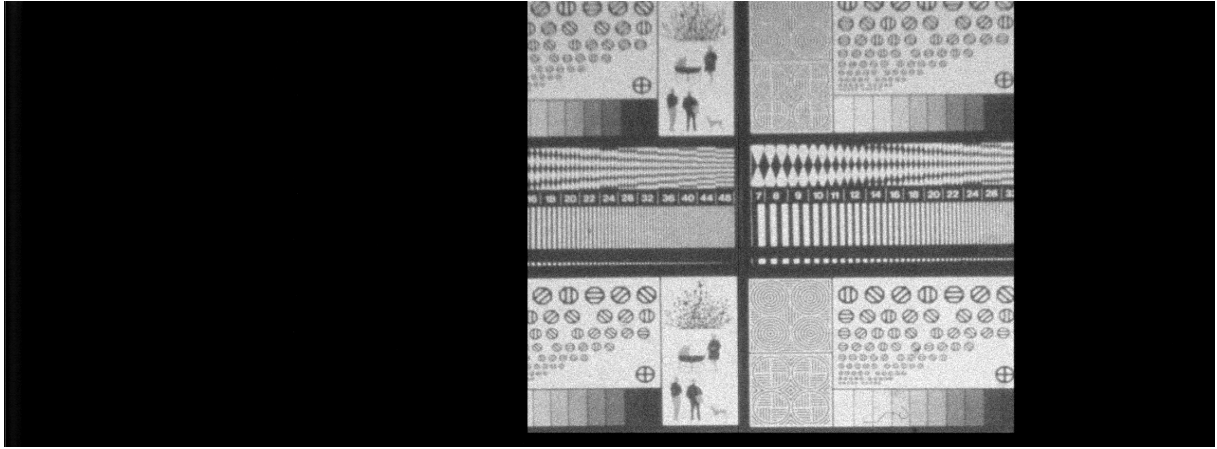

Figure 3: The image format used to make measurements of clock induced charge with electron multiplying CCDs. The right-hand section shows the image area and the left-hand section shows the serial overscan region. Parallel clock induced charge only appears in the image area combined with serial clock induced charge whereas only serial clock induced charge can be present in the overscan area.

The image from the parallel (image) register is seen on the right while on the left is the overscan region which cannot have any contribution from the parallel register. Any clock induced charge seen in the left-hand section can only be due to generation in the serial register while clock induced charge seen in the right hand area will be the sum of both parallel and serial contributions. The parallel area will be affected by clock induced charge as well as by dark signal. It is important to be certain that the dark signal is well understood and, preferably, reduced to negligible levels by deep cooling. When an electron multiplying CCD is operated at high pixel rates significant power is dissipated in the device which increases its temperature and consequently its dark signal level. The temperature monitoring is usually carried out in the metal block on which the device is mounted so it is quite possible for the true CCD operating temperature to be significantly above the indicated temperature, particularly when operating at the highest frequencies. It is a very important that once the operating temperature is stable the user confirms that the dark current is indeed negligible. It may mean that the device has to be operated at a significantly lower temperature than normally appropriate. It is also the case that when working at levels that may be as low as one photon in tens of thousands of pixels per frame that light leakage is eliminated. The sensitivity of a back illuminated CCD is very high and very few vacuum dewars are totally light proof. Great care must be taken to ensure that clock induced charge measurements are not contaminated by spurious light leakage. With our own design, light leakage through the body of the printed circuit head board is detectable, and so needs to be carefully masked off.

The first stage in quantifying clock induced charge must be to establish the electron multiplying gain of the system. This procedure has been described already by Tubbs[5]. The method involves assembling the histogram of data values from a large number of frames where the signal never exceeds one photon per pixel per frame. This is then plotted as the natural logarithm of the frequency of each data number as a function of data number (figure 4). A straight line is fitted to the high signal end of the curve and the inverse of the slope of this line gives the gain in electrons per data number.

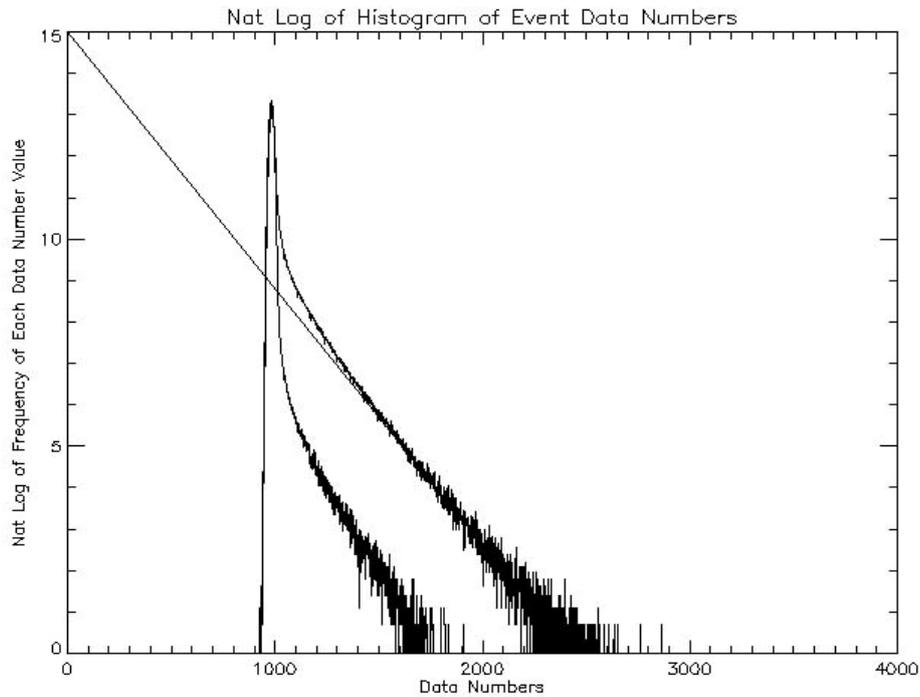

Figure 4: The data histogram as used to establish EMCCD gain. The upper curve is the histogram from a weakly illuminated image while the lower curve is the histogram from the overscan region.

Figure 4 shows two histograms, the upper taken from the parallel register (image area) with low signal levels while the lower one is taken from the overscan region and therefore with an assured zero signal level. The figure shows that the slope of the histogram in the overscan region where all the events are generated in the serial register is significantly steeper than that from the image area. The slope in the overscan area histogram corresponds to a gain that is approximately the square root of the gain is measured in the image area. This is because the clock induced charge in the serial register occurs randomly within the multiplication part of the serial register. On average a serial clock induced electron is multiplied with only half the number of transfers so the average gain is the square root of the gain achieved with the full number of transfers

The characterisation of the clock induced charge in the serial register is important. Firstly, it is important to remember that no matter how high the gain might be in the multiplication register the most likely gain level for a specific photon is still unity. For photon counting operation a threshold level must be chosen such that all events above this threshold are assumed to be genuine photons. Setting the threshold to a higher level guarantees that there is minimal confusion with read noise or serial clock induced charge, but at the expense of discarding significant numbers of events that are below the threshold and yet are indeed genuine photons. However if the threshold is set to a level which allows contamination by the tail of the Gaussian noise distribution then some events will be spurious. When clock induced charge in the serial register is considered then this distorts the tail of the event distribution (see figure 4) so that even if read noise events are discarded by careful selection of the threshold it is still possible for clock induced events to contribute significantly. Again, too high a threshold will guarantee the absence of clock induced charge events but only at the expense of removing significant numbers of genuine photon events.

At any particular signal level the user must determine what is an acceptable background clock induced charge level. The gain employed, and the clock induced charge level encountered allow an estimate of the number of genuine photon events rejected for a particular threshold together with an estimate of the contamination by clock induced charge events. It also allows an estimate of the true detective quantum efficiency since any threshold will inevitably lead to the rejection of genuine photon events which are too low in amplitude to exceed the threshold.

# 7. CLOCK INDUCED AVALANCHE

It might be imagined that at very low signal levels it is possible to use very high gain values with electron multiplying CCDs. The difficulty arises when working in high gain is that as the charge packet is transferred through the multiplication register the accumulated charge level can be large enough to affect the multiplication statistics within individual serial pixels. This can cause the multiplication process to be somewhat unstable and lead to an apparent increase in gain, the further along the serial multiplication register the charge progresses. This is not a simple serial register saturation problem since the signal levels involved are still well below register full well.

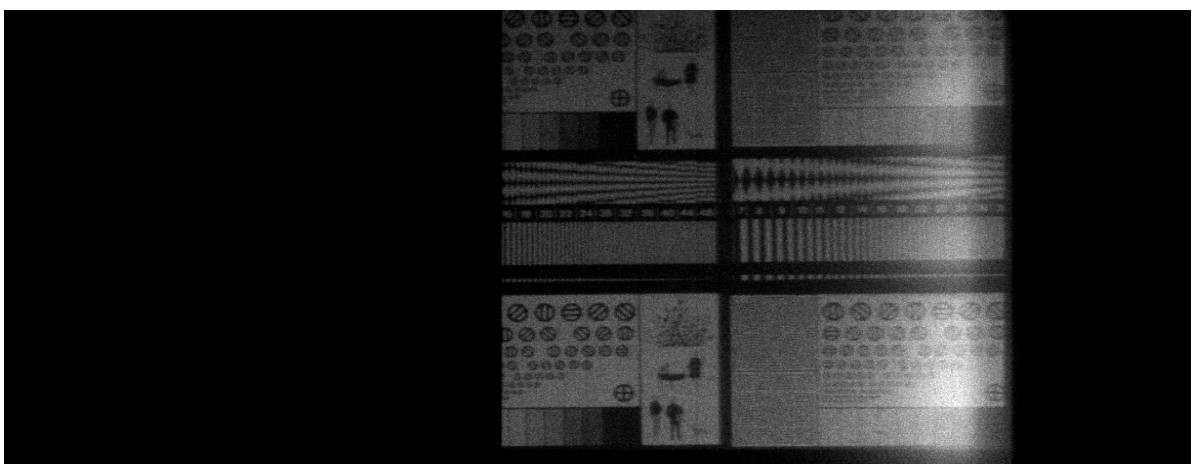

Figure 5: the effect of imaging with an electron multiplying CCD is set to an excessively high gain level. This is the average of many very low light level images with a gain of about 10,000. The gain is considerably higher on the right hand side of the image because of clock induced avalanche multiplication.

Nevertheless it is entirely visible on an image such as figure 5 which is identical to figure 3 apart from the fact that the electron multiplying CCD 201 is being operated at a gain in the region of 10,000. It is important to ensure that whatever gain is chosen does not risk this effect becoming important. The effect seems to be some kind of avalanche effect due to a combination of the high potentials needed for very high gains plus the accumulated signal level in the multiplication register pixels. This effect markedly affects gain linearity in those regions where this is significant.

# 8. EMCCD AGEING ISSUES

Standard CCDs are well known to be remarkably insensitive to light overload. Indeed it is one reason that CCDs are used as widely as they are. E2V Technologies Ltd. EMCCDs, however suffer from ageing that in extreme cases will cause the device to fail. It is worth noting that Texas Instruments EMCCDs are very much less affected by this[10]. Ageing is seen as a drop in multiplication gain as the sensor is run. The drop in gain can be corrected for by increasing the amplitude of the high-voltage clock signal by up to 4 or 5 volts at which point the device stops transferring charge properly. There appear to be two aspects of the ageing, a short-term one that dominates during the first ~1000 hours of operation, and a long-term one that characterises subsequent ageing. According to E2V Technologies[11] the short-term ageing depends significantly on signal level. It doubles with a gain increase from x100 to x1000. E2V specify a maximum gain of x1000 for all their devices, and with the gains needed for photon counting (typically x2000), short-term ageing must be expected to be more severe. At gains below x2, no ageing is observed. The long-term ageing, however, appears to be relatively insensitive to signal level. With careful use, and by avoiding over-illumination and/or gain settings that gives output register signal levels close to saturation, operational lifetimes of tens of thousands of hours may be obtained (see Figure 6). However, with a saturated image area (above the image area full well specification of ~250,000 electrons/pixel/frame and the CCD operated at a gain of x1000, the device will fail typically in only 5 hours.

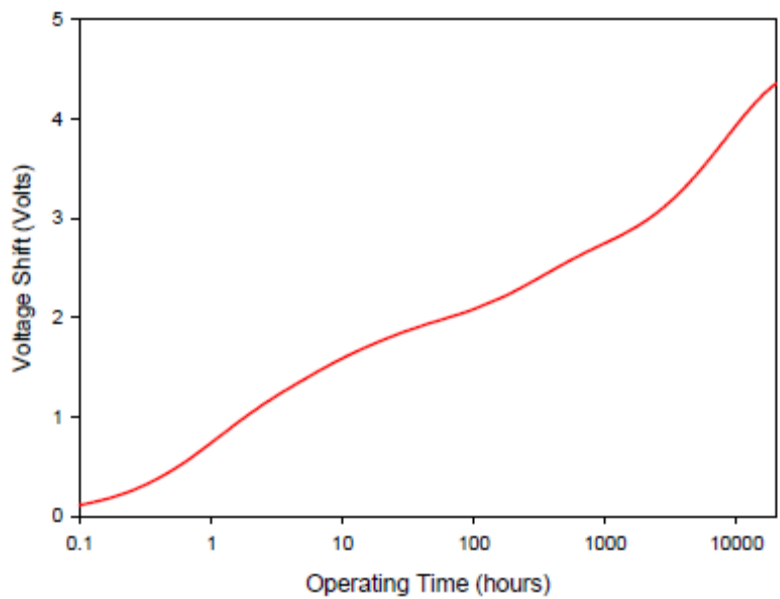

Figure 6: The ageing characteristics of an E2V Technologies EMCCD operated with a gain of x1000 and an image illumination level of ~300 electrons per pixel per frame at 11 MHz pixel rate. At this level, a device lifetime in excess of 20,000 hours (over 3 years of uninterrupted operation). See reference [11].

## 9. APPLICATION EXAMPLE 1: LUCKY IMAGING

Lucky Imaging is the name given to the technique of taking images on ground-based telescope fast enough to freeze the motion is due to atmospheric turbulence [6,7]. Provided there is a relatively bright compact reference object within the field of view, images are selected on the basis of their sharpness then shifted and added to give a summed image. Different percentages may be selected and, for the right size of telescope on a good observing site (for example, a 2.5 m diameter telescope working in I band on la Palma), near-diffraction limited images are obtained with 10-30% of the observations, depending on conditions. Even larger percentages may be used at the expense of somewhat degraded resolution.

We have developed a camera that uses four CCD 201 EMCCDs, each with 1024 x 1024 pixels of 13 μ square. A contiguous focal plane of 4096 x 1024 pixels are reimaged using steerable mirrors on to the four separated detectors. The detectors are run fully synchronised at 26 MHz pixel rate and produce 208 GB of data per second. Lucky Imaging requires a reference object within the field of view that is bright enough to allow image quality to be assessed reliably. In practice this calls for an object that gives in excess of 100 detected photons per frame. Even at the high magnification that is necessary in order to realise near diffraction limited image resolution there are regions of the reference star images which are well above the one photon per pixel per frame limit above which photon counting becomes difficult. For this reason the selection is carried out using analogue images around the reference star while the rest of the field is thresholded to give the optimum signal-to-noise on the science targets which may be very much fainter than the reference object. When working at rather low photon rates the thresholding not only doubles the equivalent quantum efficiency by removing the variance on the event amplitudes but also eliminates detector non-uniformities due to electronically-induced settling to give a completely uniform background. This approach gives very good quality images such as the example shown in figure 7.

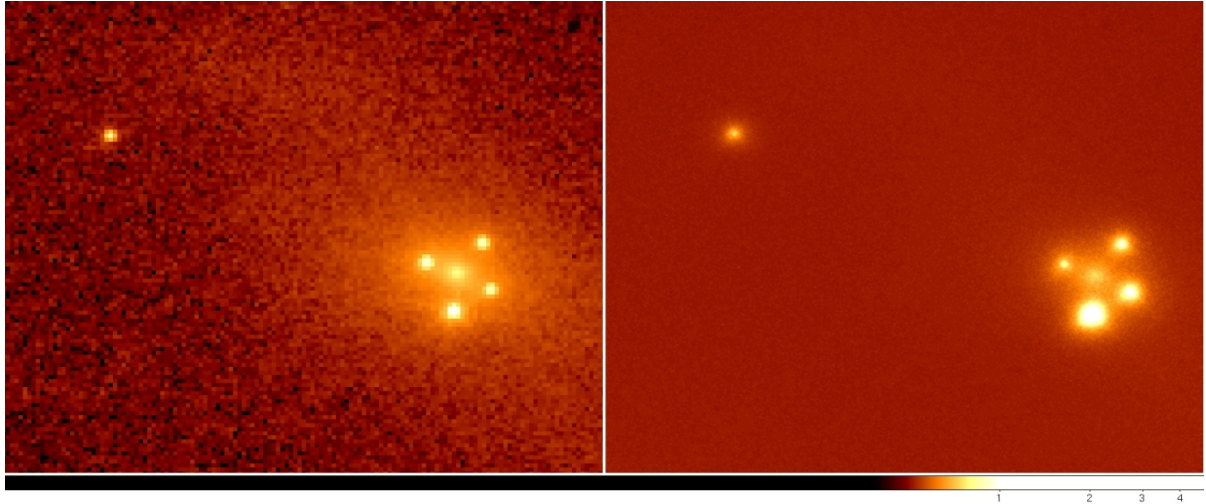

Figure 7: Images taken of the Einstein cross, which shows four gravitationally lensed images of a distant quasar magnified by the core of a relatively nearby bright galaxy. The left-hand image was taken with the Hubble Space Telescope Advanced Camera and the right hand image taken with the Cambridge LuckyCam attached to the NOT 2.5 m telescope on La Palma.

## 10. APPLICATION EXAMPLE 2 CURVATURE WAVEFRONT SENSING[12]

Adaptive optic systems are used to correct wavefront errors in the imaging systems where the atmospheric turbulence smears out astronomical images with ground-based telescopes. Most adaptive optics systems use Shack-Hartmann sensors with a lenslet array to create a grid of star images each from a different portion of the telescope pupil. For high order correction work this is often a good approach but, for lower order correction, curvature sensors have been shown by Racine (2006) to be much more efficient and allow operation on much fainter reference stars. The usual approach to the design of curvature sensor detectors has been to use a number of avalanche photodiode detectors in conjunction with specially shaped lenses to gather the light from different parts of the pupil plane. Guyon et al[9] has shown that by imaging the wavefront intensities on either side of the pupil plane at four different positions it is possible to determine the wavefront errors very efficiently.

We have designed an optical arrangement that splits the light from a reference star into four beams so that all four planes may be imaged simultaneously on to a photon counting EMCCD. By spreading the light from each pupil image over a diameter of ~200 pixels of a fast readout camera system, even relatively bright reference stars will not exceed the one photon per pixel per frame threshold that is determinant of successful photon counting performance. Once each image is in the computer, regions of the pupil images may be combined spatially and temporally to allow determination of the wavefront correction needed for the low order adaptive optics system.

## 11. APPLICATION 3: HIGH RESOLUTION FAINT OBJECT SPECTROSCOPY

Lucky Imaging allows high resolution images to be obtained with ground based 2.5 m diameter class telescopes. Lucky Imaging systems when combined with a low order curvature sensor-based adaptive optics subsystem as described in the previous section allow high resolution images to be obtained on much larger telescopes using reference stars as faint as 18.5 magnitude. These high resolution images may be projected onto an integral field fibre bundle so that each fibre only samples a small part (~0.1 arc second diameter) of the image. The outputs of the fibres are set in a line that forms the slit input to a spectrograph and spectroscopic data are recorded at high resolution. The reference star is imaged by a similar fibre bundle and guides the image selection and positioning of the accumulated image by using an undispersed image of the fibre output. Both science detector and wavefront sensor need to be based on EMCCD technologies in order to work efficiently at the very low light levels and high frame rates that are used. In this way near diffraction limited spectroscopy may be carried out on large ground-based telescopes.

## 12. CONCLUSIONS

Electron multiplying CCD detectors represent a major breakthrough in CCD technology. They offer astronomers a number of interesting new opportunities for the design and operation of astronomical instruments, both in direct imaging and spectroscopy. In photon counting applications the loss in effective quantum efficiency that the multiplication register otherwise imposes may be eliminated. With careful design, clock induced charge effects may be reduced to negligible levels. Avoiding excessive illumination will allow many thousands of hours of operation. EMCCDs enable the development of high speed imaging systems that will give insight into many rapidly varying astronomical objects. In addition, they offer the possibility of achieving routinely near diffraction limited imaging and spectroscopy on large (8-10 m class) ground-based telescopes and so giving information on much smaller angular scales then is possible with present and planned space based instruments. They also have the potential to deliver even higher angular resolution on the next generation of larger ground-based telescopes.

## REFERENCES

[1] P. Jerram et al, Proc SPIE vol 306, p178-186, (2001)
[2] G. Basden, C. A. Haniff and C. D. Mackay, (2003) "Photon Counting Strategies with low light level CCDs", Mon. Not R. astro. Soc, vol. 345, 985-991.
[3] E2V Technologies Ltd, "Lowlight Technical Note 4: Dark Signal and Clock Induced Charge in L3 Vision CCD Sensors" (2004)
[4] E2V Technologies Ltd, "Low Light Technical Note 5,:An Overview of the Ageing Characteristics of L3VisionTM Sensors" (2006)
[5] Tubbs, R., "Lucky Exposures: Diffraction limited astronomical imaging through the atmosphere", Ph.D. thesis, University of Cambridge (2003). See: http://www.mrao.cam.ac.uk/telescopes/coast/theses/rnt/node1.html
[6] Baldwin, J.E, Tubbs, R.N., Mackay, C.D., et al, (2001), Diffraction-limited 800nm imaging with the 2.56m Nordic Optical Telescope, Ast & Astrophys, vol. 368, L1-4.
[7] Baldwin, J.E, Warner, P.J., and Mackay, C.D., "The Point Spread Function in Lucky Imaging and Variations in Seeing on Short Timescales", (2008), Astron & Astrophys, vol. 480, p589
[8] R. Racine, "The Strehl Efficiency of Adaptive Optics Systems", (2006), Pub. Astr. Soc Pacific, vol 118, p1066-1075.
[9] O. Guyon, et al, Improving the Sensitivity of Astronomical Curvature Wavefront Sensor Using Dual Stroke Curvature, PASP, 120, 655 (2008).
[10] Andor, private communication, (2009).
[11] E2V Technologies Ltd," Low Light Technical Note 5:An Overview of the Ageing Characteristics of L3VisionTM Sensors" (2009).
[12] Craig Mackay, Tim D. Staley, David King, Frank Suess and Keith Weller (2010) "High-resolution imaging and spectroscopy in the visible from large ground-based telescopes with natural guide stars.", Proc SPIE 7735.